\documentclass[namedreferences]{solarphysics}
\pdfoutput=1
\usepackage[optionalrh,showbiblabels]{spr-sola-addons} 
\usepackage{hyperref}
\usepackage{graphicx}        
\usepackage{amssymb}        
\usepackage{color}           
\usepackage{breakurl}        

\renewcommand{\vec}[1]{{\mathbfit #1}}
\newcommand{\deriv}[2]{\frac{{\mathrm d} #1}{{\mathrm d} #2}}

\newcommand{\uvec}[1]{ \hat{\mathbf #1} }
\newcommand{\pder}[2]{ \frac{\partial #1}{\partial #2} }
\newcommand{\grad}{ {\bf \nabla } }
\newcommand{\curl}{ {\bf \nabla} \times}

\newcommand{\ap}{ \vec A_p}

\newcommand{\bb}{\vec B}

\chardef\us=`\_

\begin{document}

\begin{article}
\begin{opening}

\title{Optimization of photospheric electric field estimates for accurate retrieval of total magnetic energy injection\\ {\it Solar Physics}}

\author[addressref=aff1,corref,email={
erkka.lumme@helsinki.fi}]{\inits{E.}\fnm{E.}~\lnm{Lumme}\orcid{0000-0003-2045-5320}}
\author[addressref=aff1]{\inits{J.}\fnm{J.}~\lnm{Pomoell}\orcid{0000-0003-1175-7124}}
\author[addressref=aff1]{\inits{E. K. J.}\fnm{E.~K.~J.}~\lnm{Kilpua}\orcid{0000-0002-4489-8073}}

\address[id=aff1]{Department of Physics, University of Helsinki, P.O. Box 64, 00014 Helsinki, Finland}

\runningauthor{E. Lumme et al.}
\runningtitle{Optimization of photospheric electric field estimates}

\begin{abstract}
Estimates of the photospheric magnetic, electric and plasma velocity fields are essential for studying the dynamics of the solar atmosphere, for example through the derivative quantities of Poynting and relative helicity flux and by using of the fields to obtain the lower boundary condition for data-driven coronal simulations. In this paper we study the performance of a data processing and electric field inversion approach that requires only high-resolution and high-cadence line-of-sight or vector magnetograms -- which we obtain from \emph{Helioseismic and Magnetic Imager} (HMI) onboard \emph{Solar Dynamics Observatory} (SDO). The approach does not require any photospheric velocity estimates, and the lacking velocity information is compensated using ad hoc assumptions. We show that the free parameters of these assumptions can be optimized to reproduce the time evolution of the total magnetic energy injection through the photosphere in NOAA AR 11158, when compared to the recent state-of-the-art estimates for this active region. However, we find that the relative magnetic helicity injection is reproduced poorly reaching at best a modest underestimation. We discuss also the effect of some of the data processing details on the results, including the masking of the noise-dominated pixels and the tracking method of the active region, both of which have not received much attention in the literature so far. In most cases the effect of these details is small, but when the optimization of the free parameters of the ad hoc assumptions is considered a consistent use of the noise mask is required. The results found in this paper imply that the data processing and electric field inversion approach that uses only the photospheric magnetic field information offers a flexible and straightforward way to obtain photospheric magnetic and electric field estimates suitable for practical applications such as coronal modeling studies. 
\end{abstract}
\end{opening}

\section{Introduction}
     \label{S-Introduction} 
     
Determination of the photospheric magnetic, electric and plasma velocity  fields is central for understanding many of the dynamical processes in the solar atmosphere. These fields are required for quantifying the transport of magnetic flux, energy and helicity from the solar interior through the photosphere to the chromosphere and corona \citep{Liu2012,Tziotziou2013,Kazachenko2015}. They also form the essential input for \textit{data-driven modeling} of the corona \citep{Wiegelmann2015,Inoue2016,Leake2017}, in which the fields are used to specify the boundary condition at the lower radial boundary of the simulations. Due to the sparsity of direct observations of many of the key quantities that characterize the coronal plasma -- most prominently the magnetic field -- data-driven simulations offer currently the most feasible and often the only way to quantify the dynamics of the corona. Consequently, data-driven modeling is the key for studying a variety of coronal phenomena, including coronal mass ejections (CMEs) (\emph{e.g.}, \citealp{Chen2011,Pagano2013}) and flares (\emph{e.g.}, \citealp{Shibata2011,Jiang2016}), and the formation of the solar wind (\emph{e.g.}, \citealp{Riley2011}).

Remote sensing measurements of the photospheric magnetic field vector and the line-of-sight (LOS) component of the plasma velocity are routinely made at high resolution by both space-based \citep{Tsuneta2008,Scherrer2012} and ground-based instruments (\emph{e.g.}, \citealp{Mickey1996,Keller2003}). However, the available data does not directly provide the photospheric plasma velocity vector and/or the electric field, as a result making it necessary to employ rather complex inversion methods to fully retrieve these fields. A variety of methods exists for this task \citep{Welsch2007,Ravindra2008,Schuck2008,Kazachenko2014,Tremblay2015}, all sharing similar basic building blocks: The photospheric velocity/electric fields are constrained to be consistent with the available time series of the magnetic field measurements and the (ideal) magnetohydrodynamic induction equation (or Faraday's law). Moreover, most of the methods employ optical flow methods (\emph{e.g.}, \citealp{Schuck2006}) to acquire a proxy for the transverse plasma velocity component, for which no direct remote sensing observations exist. Some of the methods additionally employ the  LOS velocity estimates from Dopplergrams (\emph{i.e.}, full-disk photospheric Doppler images) as an additional constraint.

In the face of the incomplete input data that inherently includes noise as well as calibration issues \citep{Welsch2013,Lagg2015,Schuck2016}, it is clear that any given inversion method can only yield results up to a certain degree of accuracy. So far the performance of the methods has been evaluated using synthetic photospheric data extracted from a magnetohydrodynamics (MHD) simulation in which the true plasma velocity and electric fields are known \citep{Abbett2004,Welsch2007}. The best-performing methods in these tests include the PDFI method of \citet{Kazachenko2014} and DAVE4VM by \citet{Schuck2008}. In addition to quantifying the degree to which the velocity and electric fields of the simulation are reproduced, the tests of the methods have focused also on the correct reproduction of the fluxes of magnetic energy (\emph{i.e.}, Poynting flux) and relative magnetic helicity through the photosphere, as both of these quantities are central for coronal dynamics. Sufficient amount of magnetic energy, and more specifically \textit{free} magnetic energy -- energy in excess of that of the potential magnetic field configuration -- is pivotal for solar eruptions \citep{Aly1991,Chen2011, Shibata2011}. Relative magnetic helicity -- the helicity in excess of the helicity of the potential magnetic field configuration -- may have a significant role at least in some solar eruptions though its significance is still under debate \citep{Demoulin2007}. For example, \citet{Tziotziou2012} found, using non-linear force-free field modeling for a large number of active regions, that eruptive active regions tend to accumulate large budgets of both free magnetic energy and signed relative helicity. On the other hand, \citet{Pariat2017} recently found, using and a set of coronal MHD simulations in an idealized settings, that a large signed relative helicity budget does not provide a reliable eruptivity proxy. Instead they found that it is the ratio between the relative helicity in the current-carrying part of the field and the total relative helicity that best differentiates eruptive from non-eruptive coronal configurations.

The accurate estimation of the Poynting and helicity fluxes is of significant practical value when the photospheric velocity/electric fields are used to derive the photospheric boundary condition for data-driven coronal modeling, in which case the boundary condition controls the injection of these quantities to the simulation domain. Such data-driven models vary in complexity ranging from simple static (non-linear) force-free field models employing only the magnetic field as the boundary condition (see \emph{e.g.}, \citealp{Wiegelmann2012}) to full time-dependent MHD models employing fully consistent photospheric boundary conditions (\emph{e.g.}, \citealp{Jiang2016}, \citealp{Leake2017}). In terms of complexity there exists a model between these extremes called the Time-dependent MagnetoFrictional method (which we henceforth abbreviate as the TMF method). The TMF method approximates the evolution of the coronal magnetic field as an interplay between two competing  processes: relaxation towards a force-free state and a constantly evolving photospheric boundary condition \citep{Yang1986,vanBallegooijen2000,Cheung2012,Pomoell2017}. Due to the simplified coronal physics the method offers a computationally inexpensive way for coronal modeling, while still having sufficient accuracy for time-dependent simulations of the formation and possibly also of the ejection of CME flux ropes \citep{Mackay2006,Gibb2014,Fisher2015,Weinzierl2016}. The TMF method employs the photospheric electric field as its only data-driven boundary condition thereby offering a direct application for the photospheric electric field estimates. In addition, coronal TMF simulations offer an additional way for evaluating the performance of the inversion methods through studying the realism of the simulation feedback to the electric field boundary condition (\emph{e.g.}, \citealp{Cheung2012, Cheung2015}).

In this paper we study how the fluxes of magnetic energy and relative helicity depend on the choice of the electric field inversion method using a data processing and electric field inversion approach that requires only high-resolution and high-cadence (vector) magnetograms as an input and does not use any photospheric velocity estimates. The approach is flexible to use, but the accuracy of the electric field estimates is restricted as the missing photospheric velocity information must be completed using ad hoc assumptions. We show that, despite these limitations, we are able to optimize the free parameters of the applied electric field inversion methods to yield accurate estimates for the total (and implicitly also the free) magnetic energy injected through the photosphere. Our study focuses on the active region NOAA AR 11158, which has been studied extensively by several authors  \citep{Schrijver2011,Cheung2012,Jing2012,Liu2012,Tziotziou2013,Kazachenko2015}. The paper is organized as follows: In Section 2 we present the data processing and electric field inversion methods used in the study and how we apply these methods to NOAA AR 11158. In Section 3 we present the main findings of our study and discuss also some of the data processing details that may affect the results, but which have not garnered much attention in the previous works. These include the masking threshold used in removing the noise-dominated pixels and the method of tracking the active region as it traverses across the solar disk. In Section 4 we discuss the implications of the results on the general applicability of our data processing and electric field inversion methods, particularly from the viewpoint of providing the photospheric boundary condition for coronal (TMF) simulations. 

\section{Data and methods}
	\label{S-data_methods}
	
In this section we first describe the methods that we use to create time series of photospheric magnetic field and electric field maps, and how we estimate the magnetic energy and relative helicity injection through the photosphere from these maps. We then present the application of these methods to NOAA AR 11158, and how we optimize our electric field estimates for accurate retrieval of total magnetic energy injection in this active region.

\subsection{Data analysis and computation tools}
	\label{S-analysis_tools_intro}
In this study we employ two tools to invert the photospheric electric field, ELECTRICIT and DAVE4VM. ELECTRIC field Inversion Toolkit (ELECTRICIT) is our newly developed software toolkit that handles all steps from downloading and processing of magnetogram data to the inversion of the photospheric electric field (explained in detail in Sections \ref{S-data_processing} and \ref{S-E-Inv}). This paper employs the entire pipeline of the toolkit. The toolkit is written in Python and its functionality relies heavily on SunPy \citep{Mumford2015} and SciPy libraries \citep{Jones2001} as well as the FISHPACK code \citep{Swarztrauber1975}, which we call from Python. Currently, the toolkit is limited to electric field inversion based on vector or LOS magnetograms only.

Differential Affine Velocity Estimator For Vector Magnetograms (DAVE4VM) is a velocity inversion code developed by \citet{Schuck2008}. When fed with a time series of photospheric vector magnetograms, the toolkit determines the photospheric velocity vector field by minimizing the vertical component of the ideal induction equation, 
\begin{equation}		\label{Eq-ideal-induct}
\pder{B_z}{t} + \nabla_h \cdot \left(B_z \vec{V}_h - V_z\bb_h \right) = 0,
\end{equation} 
in the neighborhood of each magnetogram pixel. Here $z$ refers to the direction perpendicular to the photosphere and the subscript $h$ corresponds to the components parallel to the photosphere in a local Cartesian system where the curvature of the solar surface is neglected. The output velocity field can be used to determine the electric field through the ideal Ohm's law:
\begin{equation}		\label{Eq-ideal_Ohms}
\vec{E} = -\vec{V} \times \bb
\end{equation}
In this paper we apply DAVE4VM using vector magnetograms processed by  ELECTRICIT (see Section \ref{S-E-Inv} for details).

\subsection{Processing of vector magnetograms}
	\label{S-data_processing}
Our photospheric data input consists of full-disk disambiguated vector magnetograms provided by the \emph{Helioseismic and Magnetic Imager} (HMI) instrument \citep{Scherrer2012} onboard the \emph{Solar Dynamics Observatory} (SDO) spacecraft \citep{Pesnell2012} described in detail by \citet{Hoeksema2014}. The magnetograms are provided with spatial resolution of 0.5'' per pixel in the plane of sky and a cadence of 720 seconds. In each pixel the magnetic field is inverted from the full Stokes vector observations made at a cadence of 135 seconds averaged over a tapered time window of 1350 seconds. We use ELECTRICIT to download these magnetograms from the Joint Science Operations Center (JSOC) under the dataset title \verb+hmi.B_720s+ using a modified version of the download functionality of the SunPy library.

As explained by \citet{Hoeksema2014} the vector magnetogram data product includes four different options for the disambiguation of the magnetic field vector. Strong field pixels (nominally pixels where $B >$ 150 Mx cm$^{-2}$) are disambiguated using the minimum energy method \citep{Metcalf1994} whereas for weak field pixels the user can decide between three disambiguation methods: the potential field acute angle method, radial acute angle method or a randomly chosen disambiguation. We choose to use the potential field acute angle method for the weak field pixels. We recognize the fact that the method can produce unphysical artifacts \citep{Liu2017}. However, the approach constrains the azimuth using a physically justifiable condition (unlike the randomly chosen disambiguation), and it is applicable over the entire solar disk (unlike the radial acute angle method that in practice yields a random disambiguation near the disk center, see \emph{e.g.}, \citealp{Gosain2013}). Since we mask the noise-dominated weak field pixels to zero before the inversion of the electric field the choice of the weak field disambiguation method is not significant for the results presented in this paper (see further discussion in Section \ref{S-Effect_of_tracking}).

The HMI vector magnetograms exhibit spurious bad pixels in regions where the Stokes inversion code fails. Since bad pixels can be recognized by exceptionally high formal errors given by the Stokes inversion module \citep{Hoeksema2014}, we use a fixed threshold for the error of the total field strength $\sigma_B = 750$ Mx/cm$^2$, above which we label each pixel as bad. This threshold recognized all bad pixels in the strong field region for the test cases we considered (see \emph{e.g.}, \citealp{Hoeksema2014}), but seems to be slightly oversensitive in the weak field region. We substitute the vector components of the bad pixels (in the HMI image basis, \citealp{Sun2013}) by the median of the good pixels in the nearest symmetric neighborhood of each bad pixel that contains at least one good pixel.

After processing the full-disk magnetograms ELECTRICIT tracks the user-specified region and creates a time series of re-projected vector magnetograms centered at the given region and projected to local Cartesian coordinates, similarly to the Space weather HMI Active Region Patches (SHARPs) \citep{Bobra2014,Hoeksema2014}. The tracking of the active regions is done similarly to SHARP data \citep{Sun2013,Hoeksema2014} with a few modifications discussed further in Appendix \ref{S-Appendix-Mods_to_SHARP_tracking}. As an additional processing step we remove the spurious temporal flips in the azimuth of the magnetic field using a slightly modified version of the method presented by \citet{Welsch2013} (see Appendix \ref{S-Appendix-Mods_disambig_flip_removal} for details).

Finally, before using the time series of processed and re-projected magnetic field vector data in the electric field inversion we mask the noise-dominated weak field pixels to zero in order to avoid introducing noise to the solution. Since our electric field inversion methods use three subsequent magnetograms at $t-\Delta t$, $t$ and $t+\Delta t$ to compute the electric field at time $t$ (Section \ref{S-E-Inv}), we demand the total magnetic field strength to be over the noise threshold in all three frames. We employ a noise threshold of $B = 250$ Mx/cm$^2$ from \citet{Kazachenko2015}, which is 2.5 times the nominal noise level $\sigma_B = 100$ Mx cm$^{-2}$ in HMI vector magnetograms \citep{Hoeksema2014}. We reproduced the determination of the noise level in the local Cartesian magnetic field components from \citet{Kazachenko2015} for our data (see Section \ref{S-Effects_of_mask_and_tracking}), and due to the use of different data processing and disambiguation methods our noise levels turned out to be smaller. However, as it will be shown in Section \ref{S-Effects_of_mask_and_tracking}, the noise threshold of the mask has a notable effect on the results, and hence, to enable optimal compatibility to the results of \citet{Kazachenko2015}, we use the same noise threshold. For DAVE4VM input we do not mask the data since DAVE4VM cannot handle input data that has large regions of uniform (zero) values. Instead, we mask the data after the velocity inversion has been performed.

\subsection{Electric field inversion}
	\label{S-E-Inv}
	
Currently, the electric field inversion within ELECTRICIT is based on a combination of the PTD part from the PTD-Doppler-FLCT-Ideal (PDFI) method \citep{Kazachenko2014} and ad hoc assumptions proposed by \citet{Cheung2012} and \citet{Cheung2015}. The PDFI approach has yielded good results in tests using synthetic data and it is also flexible allowing an inversion in the case when only part of the required input data is available (as in our case). At the core of the method is the decomposition of the electric field into inductive $\vec{E}_I$ and non-inductive $-\grad{\psi}$ components:
	\begin{equation}		\label{Eq-ind_nind_decomp}
		\vec{E} = \vec{E}_I - \grad{\psi}.
	\end{equation}	  
The inductive component is completely constrained by a times series of vector magnetograms (from which $\partial \vec{B}/\partial t$ can be estimated) and Faraday's law: 
	\begin{equation}		\label{Eq-F_law_ind}
		\curl{\vec{E}_I} = -\pder{\vec{B}}{t},
	\end{equation}	 
which in the PDFI method is uncurled using a Poloidal-Toroidal Depomposition (PTD). We solve the inductive component similarly to \citet{Kazachenko2014} with some modifications to the applied numerical methods (see Appendix \ref{S-Appendix-num_impl_of_E_inv}). Our electric field estimates are exactly consistent with Faraday's law and the evolution of $B_z$ in our time series of magnetograms. 

The inversion of the non-inductive component is more challenging. In PDFI method this is done using the ideal Ohm's law (Eq. \ref{Eq-ideal_Ohms}) and three constraints from measurements: Dopplergram velocities near PILs, optical flow velocities obtained using the FLCT method \citep{Fisher2008} and the ideal constraint $\vec{E} \cdot \bb = 0$ (giving the Doppler-FLCT-Ideal part of the method title). Although this method provides the state-of-the-art inversion results in tests, we use a more simplified approach because velocity data processing including the calibration of Dopplergrams and determination of FLCT velocity is yet to be implemented in the ELECTRICIT pipeline. Instead, we use one of the following three (ad hoc) assumptions to determine the non-inductive potential $\psi$:
\begin{enumerate}
	\setcounter{enumi}{-1}
	\item $\grad{\psi} = 0$ \label{Eq-Ass0}
	\item $\nabla_h^2 \psi = -\nabla_h \cdot \vec{E}_h = -\Omega B_z$ \label{Eq-Ass1}
	\item $\nabla_h^2 \psi = - U j_z = -U(\curl{\vec{\bb}}) \cdot \uvec{z}$, \label{Eq-Ass2}
\end{enumerate}
where $\Omega$ and $U$ are freely chosen constants. The zeroth assumption simply sets the non-inductive component to zero. \citet{Fisher2010} studied the performance of such an inversion in reproducing the electric field from simulated photospheric data, and found that the assumption reproduces the known electric field poorly, \emph{i.e.}, the contribution from $-\grad \psi$ is significant and cannot be approximated to zero. Nevertheless, we retain this assumption in this study as a point of reference and due to the fact that there are simulation studies which have reported realistic coronal evolution despite the use of the assumption in deriving the photospheric boundary condition (\emph{e.g.}, \citealp{Gibb2014}). Thus it is useful to quantify how large the deviations in the magnetic energy and relative helicity fluxes are when this assumption is used in inverting the electric field from real magnetograms. Assumption \ref{Eq-Ass1} was conceived by \citet{Cheung2012} who found that Assumption \ref{Eq-Ass0} did not produce enough free magnetic energy for their data-driven TMF simulation of flux rope ejections. Assumption \ref{Eq-Ass1} was introduced as an ad hoc method to increase the free energy budget to more realistic values without making the electric field inversion too complicated. 
As noted by the authors, the assumption imposes uniform vortical motions to the photospheric plasma. Indeed, using the ideal Ohm's law one can show that Assumption \ref{Eq-Ass1} is obtained for the case of a vertical axisymmetric flux tube rotating at a constant angular velocity $\omega = -\Omega/2$. Assumption \ref{Eq-Ass2} was presented by \citet{Cheung2015} for their TMF simulation study of recurrent homologous helical jets. They proposed that the jets were driven by an emergence of a twisted current-carrying magnetic structure, and consequently, Assumption \ref{Eq-Ass2} is obtained for the case of a uniform emergence ($\vec{V} = U\uvec{z}$) of an axisymmetric flux tube, the axis of which is parallel to the $z$ direction, and which is invariant along its axis. In ELECTRICIT we solve the Poisson equations of Assumptions 1 and 2 using the same numerical machinery as in solving the inductive component (see Appendix \ref{S-Appendix-num_impl_of_E_inv} for details). Determination of the optimal values for the free parameters $\Omega$ and $U$ is discussed in Section \ref{S-Optimization_met}.

In the DAVE4VM-based inversion we calculate the electric field from the output velocity field of DAVE4VM using the ideal Ohm's law (Eq. \ref{Eq-ideal_Ohms}). For each time $t$ the input to DAVE4VM consists of a vector magnetogram at time $t$, estimates for the horizontal derivatives of the vector components at that time (estimated using the 5-point optimized derivatives as in \citealp{Schuck2008}), and an estimate for $\partial B_z/\partial t$ calculated from the magnetograms at time $t + \Delta t$ and $t - \Delta t$ using a central difference scheme. Defining the input in this way gives the output velocity field estimate also at time $t$. The window size in DAVE4VM is $19 \times 19$ pixels, the same as used by \citet{Liu2012}, and it maximizes the Pearson correlation between the two terms of the induction equation for the vector magnetogram data that we use in this study (Section \ref{S-impl_of_methods_to_11158}). The correlation is not very good ($\sim$0.6) due to the large noise levels in the input vector magnetograms (Y. Liu, 2017, private communication), and thus DAVE4VM-based electric fields fulfill Faraday's law poorly. Consequent discrepancies introduced to the time evolution of $B_z$ prevent using these electric fields directly as the data-driven boundary condition of coronal simulations (as already discussed by \citealp{Schuck2008}).

\subsection{Calculation of magnetic energy and relative helicity injections through the photosphere}
	\label{S-E_and_H_calc}
The magnetic energy injected through the photosphere $E_m$ to the upper solar atmosphere can be calculated by integrating the Poynting flux over each magnetogram and integrating this total energy flux in time (\emph{e.g.}, \citealp{Kazachenko2015}):
	\begin{equation} \label{Eq-E_injection}
		E_m(t) = \int_0^t dt' \ \deriv{E_m}{t} = \int_0^t dt' \int \ dA \ S_z = \frac{1}{\mu_0} \int_0^t dt' \int \ dA \ (\vec{E} \times \bb) \cdot \uvec{z}   
	\end{equation}

The relative helicity $H_R$ injected through the photosphere can be calculated similarly \citep{Berger1984,Demoulin2007}:
	\begin{eqnarray} 
	H_R(t) &=& \int_0^t dt' \ \deriv{H_R}{t} = -2 \int_0^t 			dt' \int \ dA \ (\ap \times \vec{E}) \cdot \uvec{z} 		\\
	&=& -2 \int_0^t dt' \int \ dA \ (A_p^x E_y - A_p^y E_x) 		\label{Eq-H_injection} 
	\end{eqnarray}
Here $\ap$ is the vector potential that corresponds to the potential field extrapolation derived from the $B_z$ component. Similarly as in \citet{Kazachenko2015} we calculate $\ap$ in ELECTRICIT using the PTD formalism for the (masked) magnetic field using exactly the same numerical machinery as in solving the inductive electric field (Appendix \ref{S-Appendix-num_impl_of_E_inv}).

\subsection{Implementation of the analysis pipeline for NOAA AR 11158}
	\label{S-impl_of_methods_to_11158}

In this paper we employ the analysis pipeline described in Sections \ref{S-analysis_tools_intro} - \ref{S-E_and_H_calc} to the active region NOAA AR 11158. The active region emerged on February 10, 2011 on the southern hemisphere of the Sun, and its strongest manifestation of activity was the X2.2 flare (onset on February 15 at 01:44 UT) and the associated halo CME \citep{Schrijver2011}. The magnetic energy and helicity budgets of this active region have been studied extensively using photospheric estimates \citep{Liu2012,Tziotziou2013,Kazachenko2015} and force-free extrapolations (\emph{e.g.}, \citealp{Jing2012,Sun2012,Aschwanden2014}). The results of both approaches have been reviewed by \citet{Kazachenko2015}. \citet{Cheung2012} did also a series of illustrative data-driven simulations for the active region using semi ad hoc electric field inversion from LOS magnetic field data as the photospheric boundary condition. They employed Assumption \ref{Eq-Ass1} from Section \ref{S-E-Inv} and tested two values for the free $\Omega$ parameter (discussed further in Section \ref{S-Discussion_and_conclusions}).

\begin{figure}   
    \centerline{\includegraphics[width=\textwidth, trim = 2cm 0.0 0.0 0.0]{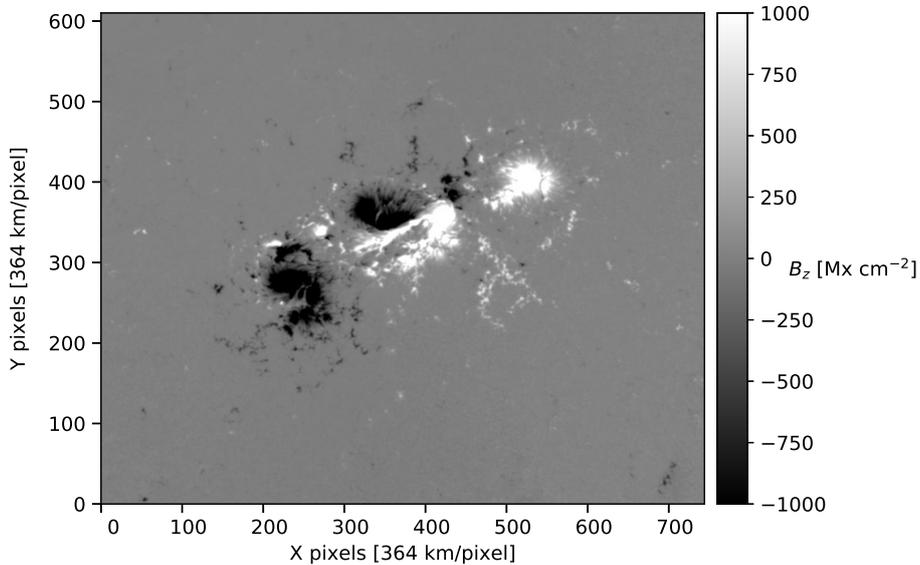}}
   \caption{Snapshot from the time series of re-projected vector magnetograms used in this paper. The plot illustrates the $B_z$ component of the magnetic field on February 15 01:36 UT when the center of the active region was at (S21,W13).} 
   \label{F-11158_examp_magnetogram}
\end{figure}

For this study we created our own time series of re-projected vector magnetograms for NOAA AR 11158 using the data processing functionality of ELECTRICIT as described in Section \ref{S-data_processing}. The time series started on February 10 00:00 UT well before the emergence of the active region when the active region was at Stonyhurst longitude E53 and lasted until February 20 00:00 UT when the center of the active region was approximately at Stonyhurst longitude W77. The region was tracked at a fixed heliographic latitude S21.0769$^{\circ}$. The re-projected vector magnetograms in the series are centered at the active region and they are projected to the Mercator projection with a projection pixel size of $0.03^{\circ} \times 0.03^{\circ}$ matching the spatial resolution of HMI full-disk vector magnetograms at the disk center. We scaled the vertical component of $B_z$ by $\cos^2 \lambda'$ -- where $\lambda'$ is the latitude of the pixel in heliographic coordinates where the center of the re-projected magnetogram patch is at the equator -- in order to account for the distortion of the pixel scale in the Mercator projection \citep{Kazachenko2015}. The size of the re-projected magnetograms is $744 \times 610$ pixels, large enough to include the entire active region. An example magnetogram from the time series is shown in Figure \ref{F-11158_examp_magnetogram}.

For the energy and helicity injection studies we inverted the photospheric electric field for each frame from February 10 14:00 UT until February 18 00:00 UT. We created three electric field time series using ELECTRICIT and each of the ad hoc assumptions \ref{Eq-Ass0} -- \ref{Eq-Ass2} and one time series from the DAVE4VM velocity inversion using the ideal Ohm's law (see Section \ref{S-E-Inv} for details). We exclude the data starting from February 18, as from this date onwards the active region was already so far at the limb that the degrading temporal stability of the azimuth disambiguation introduces spurious signals to the electric fields inverted using ELECTRICIT (similar to the ''halos'' detected by \citealp{Weinzierl2016}).

\subsection{Optimization of the free parameters of the electric field inversion}
	\label{S-Optimization_met}
	
We optimize the free parameters of our electric field inversion ($\Omega$ or $U$, Section \ref{S-E-Inv}) so that the optimal values reproduce as close as possible the time evolution of the total magnetic energy injected through the photosphere $E_m(t)$ (Section \ref{S-E_and_H_calc}, Eq. \ref{Eq-E_injection}) as provided by the reference estimate (explained below). We focus on the optimization of the magnetic energy, because, as discussed in the introduction, the (free) magnetic energy injected through the photosphere is central for the initiation of solar eruptions. We consider the total magnetic energy instead of the free part of the energy for simplicity and because, in our case, the optimization of the total magnetic energy injection in fact \emph{includes} the optimization of the free energy. It can be shown that modifying the non-inductive component does not affect the injection of magnetic energy to the potential field $E_m^{pot}(t)$ and thus, by optimizing the total magnetic energy through the non-inductive component $E_m(t) = E_m^{pot}(t) + E_m^{free}(t)$ we actually simultaneously optimize the free energy part $E_m^{free}(t)$. Intuitively this argument seems obvious because the potential magnetic field configuration and the energy within it are determined solely by the photospheric distribution of $B_z$, which is in no way affected by the choice of $-\nabla \psi$. This applies also when the electric field time series is used to drive a TMF simulation, since it is the inductive component $\vec{E}_I$ (independent from the choice of $-\nabla \psi$) that determines the evolution of the photospheric $B_z$, and thus also the evolution of the potential field. However, this result can be shown also more formally using the decomposition of the 
the vertical Poynting flux into potential and free energy contributions \citep{Welsch2006}:
\begin{equation}
	\label{Eq-Sz_pot_and_free}
	S_z = S_z^{pot} + S_z^{free} = \frac{1}{\mu_0} \left[\vec{E} \times \vec{B}^{pot} + \vec{E} \times \left(\vec{B}-\vec{B}^{pot}\right)\right] \cdot \uvec{z},
\end{equation}
where $\vec{B}^{pot}$ corresponds to the magnetic field of the potential field in the photosphere: 
\begin{equation}
	\label{Eq-B_h_pot}
	\vec{B}^{pot} = -\grad{\chi} \big|_{z=0}
\end{equation}
Here $\chi$ is the scalar potential that fulfills the Laplace equation $\nabla^2 \chi = 0$ in the corona with the Neumann boundary condition $\partial \chi/\partial z|_{z=0} = -B_z$ at the photosphere. \citet{Welsch2006} showed that the flux of potential-field energy $S_z^{pot}$ in the decomposition above gives the photospheric energy injection rate to the potential part of the coronal magnetic field $d E_m^{pot}/dt$. Using similar arguments it is straightforward to show that $d E_m^{pot}/dt$ is independent from the choice of the non-inductive component:
\begin{eqnarray}
	\label{Eq-dEm_dt_pot_full}
\deriv{E_m^{pot}}{t} &=& \int dA \ S_z^{pot} = \frac{1}{\mu_0} \int dA \ \left(\vec{E} \times \vec{B}^{pot}\right) \cdot \uvec{z} \\
&=& \frac{1}{\mu_0} \int dA \ \left[\left(\vec{E}_I - \grad \psi \right) \times \left(-\grad{\chi}\right)\right] \cdot \uvec{z} \\
&=& \frac{1}{\mu_0} \int dA \ \left[\vec{E}_I \times \left(-\grad{\chi}\right)  - \grad \psi \times \left(-\grad{\chi}\right)\right] \cdot \uvec{z}  \\
&=& \frac{1}{\mu_0} \int dA \ S_z^{pot,I}  +  \frac{1}{\mu_0} \int dA \  \left(\grad \psi \times \grad{\chi}\right) \cdot \uvec{z} 
\end{eqnarray}
Now the latter term that depends on the non-inductive component is zero:
\begin{eqnarray}
	\label{Eq-dEm_dt_pot_NI}
\int dA \ \left(\grad \psi \times \grad{\chi}\right) \cdot \uvec{z} &=& \int dA \ \left[\curl{\left(\psi \grad{\chi}\right)} - \psi \curl{\grad{\chi}}\right] \cdot \uvec{z} \\
&=& \int dA \ \left[\curl{\left(\psi \grad{\chi}\right)}\right] \cdot \uvec{z} = \oint \psi \grad{\chi} \cdot d\vec{l} \\
&=& \oint \psi \frac{\partial \chi}{\partial s} \ ds = \oint \psi B_s^{pot} \ ds = 0.
\end{eqnarray}
The final line integral over the boundary of the photospheric region vanishes if we assume that the fields are sufficiently localized and we can approximate the $\psi B_s^{pot}$ to be small at the boundary \citep{Welsch2006}. Moreover, depending on the method used to compute the potential field, $B_s^{pot} = 0$ at the boundary may apply by definition (\emph{e.g.}, \citealp{Seehafer1978}). 

In order to optimize the total (and free) magnetic energy $E_m(t)$, an a priori estimate for it is required. For our NOAA AR 11158 dataset (Section \ref{S-impl_of_methods_to_11158}) used in this paper, we extract a reference estimate for the magnetic energy injection $E_m(t)$ from the dataset of \citet{Kazachenko2015}\defcitealias{Kazachenko2015}{K2015} (hereafter \citetalias{Kazachenko2015}). \citetalias{Kazachenko2015} used the PDFI electric field inversion method, which -- as discussed in the introduction and Section \ref{S-E-Inv} -- is one of the best-performing inversion methods when tested against simulation data in which the true electric field is known. Therefore, in this work, we consider it as the state-of-the-art reference, but any other estimate could be used as well. We downloaded the dataset of \citetalias{Kazachenko2015} from \url{http://cgem.ssl.berkeley.edu/} and calculated $E_m(t)$ using the Poynting flux maps and the mask provided in the dataset. We then created an ensemble of electric field inversion time series using ELECTRICIT (as explained in Section \ref{S-impl_of_methods_to_11158}) and manually searched for optimal values of $\Omega$ and $U$ that visually matched best the time evolution of $E_m(t)$ extracted from the \citetalias{Kazachenko2015} data. After finding the visually best-matching set of values we optimized $U$ and $\Omega$ further by minimizing the root-mean-square error (RMSE) between $E_m(t)$ from \citetalias{Kazachenko2015} and our data.	

\section{Results}
	\label{S-Results}
	
\subsection{Injection of magnetic energy}
	\label{S-energy_inj}

As detailed in Section \ref{S-Optimization_met} the free parameters of our electric field inversion methods ($\Omega$ and $U$, Section \ref{S-E-Inv}) were optimized so that the time evolution of the total magnetic energy injection through the photosphere $E_m(t)$ follows as closely as possible the reference result of \citetalias{Kazachenko2015}. The optimal values searched in this were found to be:
\begin{equation}
\left\{\begin{array}{ll}
\Omega = (21 \pm 0.5)/256 \times 2\pi \ \textrm{day}^{-1} \approx (0.082 \pm 0.002) \times 2\pi \ \textrm{day}^{-1} \\
U = (42 \pm 0.5) \ \textrm{m}\,\textrm{s}^{-1},
\end{array}\right.
\end{equation}
where the error bars correspond to half of the grid spacings of $\Omega$ and $U$ discrete interval used for the final minimization of RMSE between our energy injection estimate and the \citetalias{Kazachenko2015} result.

\begin{figure}   
    \centerline{\includegraphics[width=\textwidth, trim = 0.5cm 0.2cm 1.9cm 0.4cm]{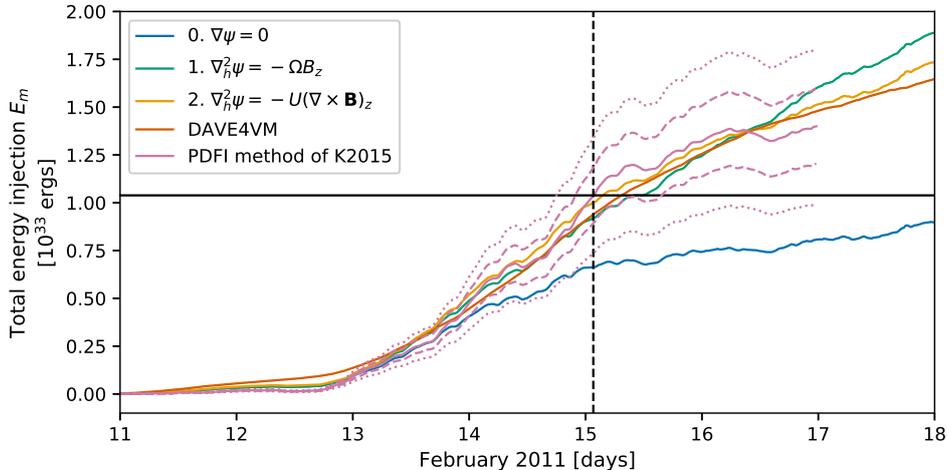}}  
   \caption{The time evolution of the total magnetic energy $E_m(t)$  injected through the photosphere in AR 11158 based on different electric field estimates. The black solid line marks the reference value from \citetalias{Kazachenko2015} on February 15 01:36 UT  (denoted by the vertical dashed line), which was the closest time before the onset of the strongest flare of the active region at 01:44 UT. The dashed and dotted purple lines represent the lower and upper error bars of the \citetalias{Kazachenko2015} dataset, respectively (see text for details).} 
   \label{F-Energy_injection}
\end{figure}

In addition to these two optimized estimates, we also created an electric field time series using Assumption \ref{Eq-Ass0} as well as the DAVE4VM velocity inversion (Sections \ref{S-E-Inv} and \ref{S-impl_of_methods_to_11158}). The evolution of the energy injection for all four estimates and the reference \citetalias{Kazachenko2015} result are plotted in Figure \ref{F-Energy_injection} for the period February 11--18, 2011. As seen from the figure, the magnetic energy evolution follows a general trend, in which first, a clear increase in the energy injection rate occurs around February 13 (which coincides with the steepest flux emergence rate of the active region). The increase continues until the X-class flare (onset on February 15 01:44 UT) after which the \citetalias{Kazachenko2015} estimate (purple curve) shows a slight decrease in the energy injection rate, which is discernible also in some of our own estimates. The estimate with vanishing non-inductive component (Assumption \ref{Eq-Ass0}, blue curve) clearly underestimates the energy injection when compared to the other estimates (\emph{e.g.}, underestimation of \citetalias{Kazachenko2015} by 36\% at the time of the flare), but reproduces quite well the overall trend of the energy injection rate before and after the flare. Assumptions 1 and 2 (green and orange curves) for the non-inductive component give a roughly similar time evolution for $E_m(t)$ when compared to each other, and they are also consistent with the DAVE4VM (red curve) and \citetalias{Kazachenko2015} results. There are, however, some obvious differences. The energy injection rate based on Assumption \ref{Eq-Ass1} does not decrease as in the other estimates which causes a clear overestimation of the energy injection from February 16 onwards. On the other hand the energy injection is underestimated between February 14 and 16. The estimate based on Assumption \ref{Eq-Ass2} is more consistent with the results of \citetalias{Kazachenko2015} underestimating its result only slightly near the time of the flare and overestimating it slightly near the end of the \citetalias{Kazachenko2015} time series (February 16 23:36 UT). There is also a good consistency between Assumption \ref{Eq-Ass2} and the DAVE4VM estimate despite the very different electric field inversion methods.

\citetalias{Kazachenko2015} provides two estimates for the relative error in the vertical Poynting flux $S_z$ and consequently also in the energy injection $E_m(t)$: the first, lower estimate is $\pm$14\% and includes only the error arising from the noise in the input vector magnetograms, whereas the second, nominal estimate $\pm$29\% includes also the method-related uncertainties of their PDFI electric field inversion method \citep{Kazachenko2014}. We consider the lower estimate more important, since the method-related uncertainties can be assumed to be larger for our electric field estimates, which are partly based on ad hoc assumptions. As illustrated in Figure \ref{F-Energy_injection} from February 14 00:00 UT until February 16 23:36 UT (the end of the \citetalias{Kazachenko2015} time series) our energy injection estimates falls well inside the bounds of the upper error estimate and almost within the lower estimate as well. More precisely, the relative difference between any of our estimates and the \citetalias{Kazachenko2015} estimate is less than 15\% over this interval. Before February 14 larger relative differences are found, but since the values of injected energy are small before this time, these corresponds to small absolute differences. Hereafter we mostly disregard this interval and discuss results only after February 14 00:00 UT.

Aside from Assumption \ref{Eq-Ass0} our energy injection estimates (Assumptions 1 and 2, and DAVE4VM) differ less than 19\% from each other from Days 14 until the end of the time series (February 18 00:00 UT).

Our results also reproduce the consistency between the \citetalias{Kazachenko2015} and DAVE4VM estimates (within both error bounds of \citetalias{Kazachenko2015}) already found by \citetalias{Kazachenko2015}, who compared their $E_m(t)$ estimate to the DAVE4VM estimates of \citet{Liu2012} and \citet{Tziotziou2013}. Our DAVE4VM estimate is completely independent from these previous studies with differences both in the input vector magnetograms fed to DAVE4VM and in the way DAVE4VM is executed (Section \ref{S-E-Inv}). From the two previous DAVE4VM studies of AR 11158 the one by \citet{Liu2012} is more consistent with ours, whereas \citet{Tziotziou2013} used a very different approach in projecting the data to a local Cartesian coordinate system. Our estimate of the energy injection overestimates the result of \citet{Liu2012} (see their Figure 14) by 5\% at reference times February 15 00:00 UT and February 16 $\sim$18:00 UT (the end of the \citealp{Liu2012} time series), and the difference is well within the error bounds $\pm$23\% reported by the authors. We overestimate the values of \citet{Tziotziou2013} (see their Figure 5) by 20\% and 54\% at the reference times above. To make a consistent comparison to these previous studies we started the time integration of the energy injection on February 12, 00:00 UT when calculating the differences above.

\subsection{Injection of relative magnetic helicity}
	\label{S-Helicity_inj_results}

\begin{figure}   
    \centerline{\includegraphics[width=\textwidth, trim = 0.5cm 0.5cm 1.9cm 0.4cm]{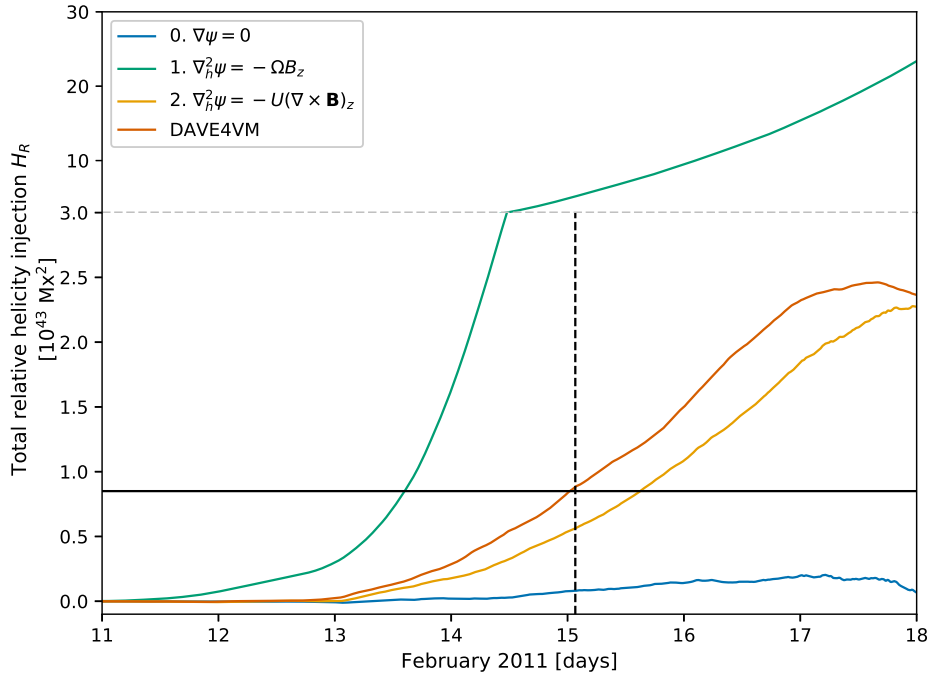}}
   \caption{The time evolution of the relative magnetic helicity $H_R(t)$ injected through the photosphere in AR 11158 based on different electric field estimates. The black dotted line marks the reference value from \citetalias{Kazachenko2015} at February 15 01:36 UT (denoted by the dashed line) near the onset time of the strongest flare of the active region.} 
   \label{F-helicity_injection}
\end{figure}

We studied also the injection of relative magnetic helicity through the photosphere $H_R(t)$ (see Section \ref{S-E_and_H_calc}) using all four of our electric field series and the optimized values of $\Omega$ and $U$ for the electric fields based on Assumptions 1 and 2 (Section \ref{S-E-Inv}). The results are plotted in Figure \ref{F-helicity_injection} with one reference data point from \citetalias{Kazachenko2015}.
As clearly seen from the figure the helicity injection has much larger differences between the different electric field estimates than in the case of energy injection (Figure \ref{F-Energy_injection}). Assumption \ref{Eq-Ass0} again clearly underestimates the result when compared to the \citetalias{Kazachenko2015} helicity injection estimate (by $\sim$90 \% at the time of the flare) as well as to the other estimates. Despite success in optimizing the magnetic energy injection for Assumptions 1 and 2 (Section \ref{S-energy_inj}) the relative helicity injection is not reproduced well for either of the assumptions. Assumption \ref{Eq-Ass1} overestimates the \citetalias{Kazachenko2015} helicity injection estimate coarsely (by a factor of 6 at the time of the flare and by a factor of 8 on February 16 00:00 UT). Assumption \ref{Eq-Ass2} matches much better the output of \citetalias{Kazachenko2015} and our DAVE4VM inversion result. The overall trends are similar, but there is now a consistent underestimation (by 34\% at the time of the flare and by 13\% on February 16 at 00:00 UT). DAVE4VM produces again results similar to \citetalias{Kazachenko2015}: our DAVE4VM estimate is only 4\% larger than the result of \citetalias{Kazachenko2015} at the time of the flare and on February 16 00:00 UT 20\% larger. Comparison to the results of previous DAVE4VM studies by \citet{Liu2012} and \citet{Tziotziou2013} on February 15 00:00 UT and February 16, 00:00 UT reveals a 45\% and a 37\% overestimation for the former (see their Figure 12), and a 8\% and a 4\% overestimation for the latter (see their Figure 5). Larger differences than the ones obtained in the case of magnetic energy injection (Section \ref{S-energy_inj}) are not surprising, considering that now, in addition to the differences in the input data and DAVE4VM setup (Section \ref{S-E-Inv}), also the computation methods of the vector potential $\ap$ are different: these authors employ FFT-based method to solve $\ap$ directly from the $B_z$ distribution whereas we employ PTD of the magnetic field solved using a finite difference method similarly to \citetalias{Kazachenko2015} (see Appendix \ref{S-Appendix-num_impl_of_E_inv} for details).

\subsection{Effects of masking and the tracking method of the active region}
	\label{S-Effects_of_mask_and_tracking}
	
\subsubsection{Effect of masking of the noise-dominated pixels}
	\label{S-Effect_of_mask}
	
As discussed in Section \ref{S-data_processing} we mask pixels to zero if $B < 250$ Mx cm$^{-2}$ in the pixel for any of the three frames that participate in the electric field inversion. The threshold was taken from \citetalias{Kazachenko2015} to ensure maximal compatibility to their results. On the other hand, according to the noise mask of the SDO/HMI vector magnetograms \citep{Hoeksema2014} the noise level of the total magnetic field strength in vector magnetograms may be as low as 60 Mx cm$^{-2}$ depending on the orbital velocity of the SDO spacecraft and the position of the observation on the solar disk. Moreover, our noise estimate for the magnetic field components ($B_x$,$B_y$,$B_z$) in the AR 11158 time series created in ELECTRICIT (Section \ref{S-impl_of_methods_to_11158}) gives maximum noise levels of $(\sigma_{B_x},\sigma_{B_y},\sigma_{B_z}) \approx (65,85,65)$ Mx cm$^{-2}$ for the time interval February 10 14:00 UT -- February 18 00:00 UT. We determined the noise levels for each magnetogram in the series by fitting a Gaussian to the weak field core ($B < 300$ Mx cm$^{-2}$, \citealp{Hoeksema2014}) of the magnetogram pixels \citep{Kazachenko2015, Welsch2013, DeForest2007}. These noise estimates imply that the ''pure noise'' magnetic field value is $B$ $\sim$125 Mx cm$^{-2}$. Therefore, when using the threshold of 250 Mx cm$^{-2}$ in masking we are removing also pixels which carry a significant signal, particularly in the relatively low-noise $B_z$ component. Thus, it is natural to ask whether changing the mask threshold affects our results, particularly the magnetic energy injection estimate used in the optimization of the free parameters $\Omega$ and $U$. In order to quantify this, we re-ran our analysis for AR 11158 using two different masking thresholds: (1) $150$ Mx cm$^{-2}$, and (2) $300$ Mx cm$^{-2}$. The first threshold corresponds to the nominal threshold used to differentiate between strong and weak field pixels in HMI vector magnetogram pipeline \citep{Liu2017}, and it is slightly above our own noise estimate of 125 Mx cm$^{-2}$. Thus, we can assume that the threshold masks at least all pure noise pixels away. The second case corresponds to the noise threshold used by \citet{Liu2012} in their DAVE4VM study of AR 11158 (Y. Liu 2017, private communication).

\begin{figure}   
    \centerline{\includegraphics[width=\textwidth]{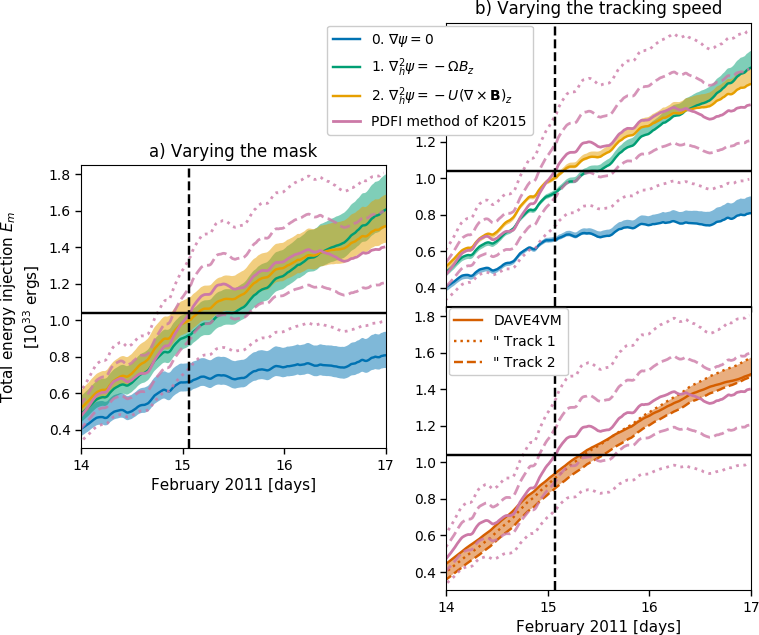}}
   \caption{Total magnetic energy $E_m(t)$ injected through the photosphere in AR 11158 and the variations arising from changes in the a) masking threshold and b) tracking speed of the active region (see text for details). The shaded region around each curve covers the variation in the energy injection due to the respective changes in the data processing. Dashed purple lines give the upper and lower error bars of the \citetalias{Kazachenko2015} result. Track 1 refers to the tracking made using the same differential rotation profile as in the SHARP data and Track 2 refers to our default tracking reinforced with coalignment. The black dotted line marks the reference value from \citetalias{Kazachenko2015} at February 15 01:36 UT (denoted by the dashed line) near the onset time of the strongest flare of the active region.} 
   \label{F-mask_and_tracking_effect}
\end{figure}

Changes in the injection of the magnetic energy with the alternative masks are illustrated in Figure \ref{F-mask_and_tracking_effect}a. We find a consistent trend that the lower the masking threshold, the larger the energy injection. When the value of the masking threshold is varied the energy injection for the three ELECTRICIT inversions (Assumption \ref{Eq-Ass0} -- 2, Section \ref{S-E-Inv}) increases/decreases as much as 21\% between February 14 and 17, the maximum median difference being 15\%. The estimate based on Assumption \ref{Eq-Ass0} is affected more than the ones based on Assumptions 1 and 2, for which both the maximum and median relative differences to the default case are $\sim$20\% and $\sim$12\%. The strongest change (increase) is found when the mask threshold of 150 Mx cm$^{-2}$ is used. Varying the mask changes the optimized estimates of the energy injection (Assumptions 1 and 2) just within the maximal error bounds of $E_m(t)$ of \citetalias{Kazachenko2015} ($\pm$29\%), but not within the lower error bounds ($\pm$14\%).

The estimate based on the DAVE4VM inversion is practically independent from the choice of the masking threshold. When the mask is varied -- or even removed entirely -- the maximum relative difference between the energy estimates is 4\% for the whole investigated interval (median relative difference being less than 2\%). Therefore, the DAVE4VM estimate is not included in Figure \ref{F-mask_and_tracking_effect}a. It is important to note that the magnetograms are not masked when fed into DAVE4VM, but the mask is applied only afterwards. Thus, the fact that DAVE4VM gives practically the same result with and without masking implies that the velocity inversion in DAVE4VM cannot retrieve any significant vertical Poynting flux signal (negative or positive) from the pixels with low signal-to-noise ratio. The same does not apply to ELECTRICIT inversions: removing the mask entirely increases the (time-integrated) energy injection even as much as 80\%. A part of this energy flux signal arises from the use of the potential acute angle disambiguation in the weak field region (Section \ref{S-data_processing}), since using the alternative methods (radial acute angle or random disambiguation) results in a smaller increase ($<40$\%) when the mask is removed entirely. Still, as already argued in Section \ref{S-data_processing} the choice of the weak field disambiguation method does not affect the results when any of the three masking thresholds considered above are applied. The reason for the spurious energy injection signal arising from the use of the potential acute angle disambiguation is likely the large-scale ''shadow-like patterns'' \citep{Liu2017} introduced to the azimuth by the method, which somehow interplay with the large-scale electric field producing a significant vertical Poynting flux signal. Nevertheless, removing the mask entirely introduces a clear increase in the energy injection regardless of the weak field disambiguation method, which implies that there are also other method-related factors in ELECTRICIT inversions that enable retrieval of a significant vertical Poynting flux signal from the noise-dominated pixels. The fact that this is not seen in DAVE4VM output implies that the signal is at least partly spurious, thus emphasizing the need for using the mask.

For the relative helicity injection we find similar trends when varying the masking thresholds as for the energy injection: \emph{i.e.}, lowering the threshold increases the injected relative helicity. Relative changes between the estimates using the alternative masking thresholds are strongest for the estimate based on Assumption \ref{Eq-Ass0} (maximum and median differences to the default are 118\% and 41\% between February 14 and 17), but the absolute changes are small and the large relative changes are simply due to the small absolute values of $H_R(t)$. For Assumptions 1 and 2 the maximum and median relative changes are 10\% and 8\%, and 19\% and 10\%. For the DAVE4VM estimate the maximum and median relative changes are: 13\% and 8\%, respectively. These changes do not alter the basic findings made in Section \ref{S-Helicity_inj_results} where the default mask was used.

\subsubsection{Effect of tracking speed of the active region}
	\label{S-Effect_of_tracking}

Another data processing parameter that likely affects the results is the speed at which the active region is tracked during its solar disk passage. Our choice for the tracking speed is based on the ''2 day lag'' differential rotation profile of \citet{Snodgrass1983}, but other options are also possible. In the SDO/HMI SHARP data product the active regions are tracked using a different profile \citep{Hoeksema2014}, which gives a 0.27 deg day$^{-1}$ ($\sim$2 km\,s$^{-1}$) faster speed for the center of AR 11158 along the line of constant latitude. On the other hand \citet{Kazachenko2015} and \citet{Welsch2013} track the active region using a coalignment procedure that minimizes the whole-frame shifts between successive frames of the magnetogram time series (in the local Cartesian coordinate system). In practice this approach results in a temporally variable tracking speed and includes also motions along the meridional direction. Due to the velocity in the ideal Ohm's law it is expected that changing the tracking speed, and thus also the velocity of the rest frame, has an effect on the electric field and thus also on the energy and relative helicity injections. 

In order to quantify how sensitive our results are to the choice of the tracking speed we re-ran our analysis for AR 11158 using two alternative tracking schemes: (1) using the differential rotation profile employed for the SHARP data product and (2) using our default differential rotation profile reinforced by coalignment which minimizes the whole-frame shifts between the successive frames. For the coalignment we used the method presented by \citet{Welsch2013} in which the coalignment is done using low-noise LOS magnetograms. For each time step we interpolated an SDO/HMI full-disk LOS magnetogram to the local Cartesian grid determined by our default tracking procedure (Section \ref{S-data_processing} and Appendix \ref{S-Appendix-Mods_to_SHARP_tracking}). Then the whole-frame shift to the previous interpolated LOS data frame was evaluated, and the local Cartesian grid was shifted so that it cancels the measured whole-frame shift. This shifted grid was then used as the final grid to which the vector (and LOS) magnetograms were interpolated at that time. For our 11158 time series the coalignment results in a tracking speed which is on average 0.28 deg day$^{-1}$ faster than our default case (with standard deviation of 0.21 deg day$^{-1}$). 

As illustrated in Figure \ref{F-mask_and_tracking_effect}b the effect of changing the tracking speed has only a small effect on the estimate of the energy injection when the three ELECTRICIT inversions (Assumptions 0 -- 2) are considered, and the changes are most of the time within the lower error bars ($\pm$14\%) of the \citetalias{Kazachenko2015} estimate. The first tracking scheme (SHARP rotation profile) produced a small increase towards the end of the time series for all three estimates, the maximum and the median relative changes being 11\% and 4\% for Assumption \ref{Eq-Ass0}, and 6\% and 3\% for Assumptions 1 and 2 between February 14 and 17. The second tracking scheme (default rotation profile and coalignment) caused only a marginal change: the maximum and the median relative differences to the default cases being at most 6\% and 2\% over all three assumptions. When the DAVE4VM estimate is considered the tracking speed has a more significant effect. Using the SHARP rotation profile the default energy injection is first underestimated but then overestimated towards the end of the time series (the maximum and median relative differences are 10\% and 3\%), whereas for the coaligned case there is a consistent decrease in the injection throughout the time series (the maximum and median relative differences: 19\% and 7\%). Despite the relatively large effect on the energy injection the DAVE4VM estimates are still most of the time within the lower error bars of \citetalias{Kazachenko2015}. The coalignment brings the DAVE4VM estimate closer to the one by \citet{Liu2012} (see Section \ref{S-energy_inj}): now the estimates differ only by 1\% near the time of the flare on February 15 00:00 UT (but still by 5\% on February 16 18:00 UT). 

The effect of the tracking speed on the estimate of relative helicity injection does not affect the main findings made in Section \ref{S-Helicity_inj_results} where the default tracking speed was used. When the tracking speed is varied the estimate based on Assumption \ref{Eq-Ass0} exhibits large relative changes ($>$100\%) due to its small absolute values, but the other estimates change by at most 8\%.

\section{Summary and discussion}
	\label{S-Discussion_and_conclusions}

In this study we examine the magnetic energy and relative helicity injection through the photosphere in NOAA AR 11158 using an ensemble of estimates for the photospheric electric field. We employ our own data processing and electric field inversion toolkit, ELECTRICIT, to create a vector magnetogram time series centered at the active region. We further create four time series of photospheric electric field maps from this data. One series is created using the DAVE4VM velocity inversion code, and the three other series are created within ELECTRICIT. In latter approach we decompose the electric field into inductive and non-inductive components. We constrain the inductive component to be exactly consistent with Faraday's law and the time evolution of $B_z$ in our vector magnetogram time series. Since constraining the non-inductive component properly from the measurements requires advanced techniques not yet implemented in our toolkit, we use three alternative ad hoc assumptions instead. This approach enables also a straightforward inversion procedure without the need of acquiring and calibrating any photospheric plasma velocity estimates. Our zeroth assumption simply neglects the non-inductive component (setting it to zero), whereas the other two constrain it by setting the horizontal divergence of the electric field proportional either to the vertical magnetic field $B_z$ or the current density $j_z$. The proportionality constants in the latter two assumptions are free parameters, and we optimize them so that the resulting electric field time series reproduce as closely as possible the time evolution of the total injection of magnetic energy from the photosphere to the corona $E_m(t)$ (computed as the area- and time-integrated vertical Poynting flux). Using analytical arguments one can show that this approach implicitly optimizes also the injection of the free magnetic energy. The reference for the optimal $E_m(t)$ is extracted from the results of \citet{Kazachenko2015} (here abbreviated as \citetalias{Kazachenko2015}), who used state-of-the-art methods to constrain the non-inductive component. 

Using the optimization scheme for the free parameters of the two semi ad hoc electric fields above we are able to reproduce the time evolution of the injected magnetic energy $E_m(t)$ well within the upper error bounds of the result of \citet{Kazachenko2015}, and almost (exceeding only by 1\%) the lower error bounds for the time period when photospheric energy injection had reached significant values (more than half of the values at the time of the strongest X-class flare of the active region). We consider the lower error bounds more important here, since they include only the error arising from the noise in the vector magnetograms, whereas the upper estimate includes also method-related uncertainties, which can be safely assumed to be much higher for our (semi ad hoc) estimates. Our DAVE4VM study of the energy injection is consistent with the results of \citet{Kazachenko2015} and with the previous DAVE4VM-based results for NOAA AR 11158 by \citet{Liu2012} within their error bounds. However, the fourth electric field time series of our ensemble, derived setting the non-inductive component to zero (our zeroth ad hoc assumption) clearly underestimates both the total (and thus also the free) magnetic energy injection. This is in line with previous simulation studies by \citet{Fisher2012} and \citet{Kazachenko2014} who showed that by assuming a vanishing non-inductive component one clearly underestimates the vertical Poynting flux. We also find that the assumption underestimates the relative helicity injection, which is again consistent with the previous results (\emph{e.g.}, \citealp{Kazachenko2014}). However, the positive sign of the total relative helicity injection is reproduced correctly. 

The underestimation of both magnetic energy and relative helicity injection implies that TMF simulations studies in which the photospheric boundary condition directly or indirectly employs Assumption \ref{Eq-Ass0}, likely underestimates the budgets of these quantities in the simulation domain (though one should keep in mind that the coronal budget of these quantities cannot be deduced solely from the photospheric fluxes, as discussed further below). Despite this implication \citet{Gibb2014} managed to successfully simulate the formation of a flux rope in NOAA AR 10977 yielding results consistent with the observed X-ray sigmoid, even though the photospheric boundary condition was based on Assumption \ref{Eq-Ass0}. However, their simulation did not include the eruption of the flux rope. Due to the central role of (free) magnetic energy and possibly also relative helicity in flux rope ejections, underestimating these quantities may be important only when the lift-off and subsequent eruption dynamics are considered. This is implied by the results of \citet{Cheung2012}, whose TMF simulation of NOAA AR 11158 gave only ''quiescent'' coronal evolution when Assumption \ref{Eq-Ass0} was used, whereas the use of Assumption \ref{Eq-Ass1} resulted in a series of flux rope ejections.

The fact that we are able to reproduce well the time evolution of the magnetic energy injection using Assumptions 1 and 2 implies that both of these semi ad hoc electric field inversion methods are flexible enough for estimation of the total (and free) magnetic energy injection. However, our approach requires a reference estimate for optimization of these assumptions through the energy injection (such as the one by \citetalias{Kazachenko2015}, used here), and for an arbitrary active region such estimates are not readily available. One straightforward way to acquire such an estimate is to use DAVE4VM: for AR 11158 it would give almost the same optimization result as the \citetalias{Kazachenko2015} estimate. If one cannot acquire an estimate for the whole time evolution of the energy injection, the reference for the optimization procedure could include an energy estimate only at some given time(s). For instance, for AR 11158 an estimate of the coronal energy content just before the strongest (X-class) flare of the active region could be used. Such an instantaneous estimate could be acquired \emph{e.g.}, from a non-linear force-free field (NLFFF) extrapolation. Though we know that the energy injection through the photosphere is not equivalent with the energy stored in the simulated coronal domain, \citet{Kazachenko2015} found that the (ordinary) NLFFF extrapolation estimates of the coronal energy content were consistent with the photospheric injection within error bounds in AR 11158.

Our optimal values of the free parameters $\Omega$ and $U$ are smaller than the values used by the original authors who proposed these semi ad hoc estimates (Section \ref{S-E-Inv}). \citet{Cheung2012} tested two non-zero values for $\Omega$, $1/8 \times 2\pi$ and $1/4 \times 2\pi$ day$^{-1}$, from which the latter was primarily used. The values are, respectively, 50\% and 200\% larger than our value $\Omega = 21/256 \times 2\pi$ day$^{-1}$. \citet{Cheung2015} tested two values for $U$, 1100 and 2200 m s$^{-1}$, from which the former was primarily used in the paper. Both values are almost two orders of magnitude larger than our value $U = 42$ m s$^{-1}$. When we run our analysis using the nominal values of the previous authors $\Omega = 1/4 \times 2\pi$ day$^{-1}$ and $U = 1100$ m s$^{-1}$ the magnetic energy and relative helicity injection are overestimated by factors 1.4 and 18 for the former case and by 9 and 15 for the latter at the time of the strongest X-class flare (when compared to the best reference values considered in this paper). Thus, our results imply that the TMF simulation of AR 11158 by \citet{Cheung2012} possibly had slightly overestimated total and free magnetic energy budgets and heavily overestimated relative helicity budget. As opposed to the TMF simulation of \citet{Cheung2012} that included the entire AR 11158, the simulation of \citet{Cheung2015} focused on homologous helical jets emanating from a small subregion of NOAA AR 11793. Thus, further conclusions about the realism of their magnetic energy and relative helicity budgets would require an additional study of the particular subregion of AR 11793. Moreover, as mentioned a few times already, we must be careful when drawing conclusions from the photospheric injections on the realism of the magnetic energy and relative helicity budgets in the simulated coronal domain. This is because part of the photospheric injection may be removed through the other boundaries of the simulation domain or may be transformed into other forms in processes within the domain. Moreover, for the special case of the magnetic energy injection in TMF simulations, not even the photospheric injections are equivalent between our estimate and the corresponding TMF simulation: we derive the photospheric Poynting flux using the horizontal components of the magnetic field from a vector magnetogram, whereas in a TMF simulation these components are determined by the magnetofrictional evolution in the coronal domain. 

A more comprehensive analysis of the relationship between the photospheric fluxes and coronal budgets would require additional (TMF) simulation studies beyond the scope of this paper. For NOAA AR 11158 we can make some further conclusions by comparing our fluxes to the simulation energy budgets reported by \citet{Cheung2012}. Their re-projected magnetograms and the resulting electric field inversion were very similar to that of this study except for the fact that they used LOS magnetograms. Using the functionality in ELECTRICIT we created an additional $B_z$ magnetogram time series from full-disk LOS magnetograms to the same grid as our nominal vector magnetogram time series (Section \ref{S-impl_of_methods_to_11158}), inverted the horizontal electric field $\vec{E}_h$ using Assumption \ref{Eq-Ass1} and $\Omega = 1/4 \times 2\pi$ day$^{-1}$, and estimated the energy injection using $\vec{B}_h$ from our nominal vector magnetogram series. Near the time of the X-class flare our photospheric injection estimate was only 13 \% larger than the total energy content reported by \citet{Cheung2012}, so at least in terms of energy the photospheric injection seems to match the coronal budget. Using this LOS magnetogram dataset we can also estimate the relative helicity injection, which is still overestimated by a factor of $\sim$16 at the time of the strongest flare of the active region. Even though the dissipation of relative helicity in the corona can be considered negligible (\emph{e.g.}, \citealp{Pariat2015}) part of the helicity injected from the photosphere can and will be removed through the boundaries of the domain, particularly during CME flux rope ejections (\emph{e.g.}, \citealp{Pariat2017}), in which case the flux rope also carries part of the helicity away from the domain. \citet{Jing2012} found in their NLFFF extrapolation study of AR 11158 that the total coronal relative helicity content was substantially lowered in tandem with the strongest M6.6 and X2.2 class flares (and the related flux rope ejections) in the active region. However, the magnitude of helicity removals (altogether $\sim \! \! 5\times 10^{42}$ Mx$^2$) is not sufficient to remove the overestimated photospheric relative helicity injection in \citet{Cheung2012} simulation implied by our results. This finding poses an interesting question whether the strong activity in the simulation (several flux rope ejections over 2 days) is partly driven by this heavy overestimation of the relative helicity budget in the corona.

Despite the relatively good optimization of the magnetic energy injection the optimal values of $\Omega$ and $U$ reproduce the injection of relative magnetic helicity through the photosphere poorly. Even with energy-optimized value of $\Omega$ Assumption \ref{Eq-Ass1} results in a heavy overestimation of the relative helicity injection. This is no surprise when one takes a closer look at the calculation of the electric field and the vector potential $\ap$. First, let's write the total flux of relative helicity arising from the non-inductive component $dH_R^{NI}/dt$ for a given time using the poloidal potential $P$ and the non-inductive potential $\psi$:
\begin{eqnarray}
\deriv{H_R^{NI}}{t} &=&-2 \int dA \ \left[\ap \times (-\grad{\psi})\right] \cdot \uvec{z} = -2 \int dA \ [(\curl{P\uvec{z}}) \times (-\grad{\psi})] \cdot \uvec{z} \nonumber \\ 
&=& 2 \int dA \ [\partial_x P \partial_x \psi + \partial_y P \partial_y \psi] 
\end{eqnarray}
From the definition of Assumption \ref{Eq-Ass1} (Section \ref{S-E-Inv}) and PTD of the magnetic field (Appendix \ref{S-Appendix-num_impl_of_E_inv}) we get:
\begin{eqnarray}
\nabla^2_h P = -B_z = \Omega^{-1} \nabla^2_h \psi 
\end{eqnarray}
Since we employ the same grid, numerical solver and homogeneous Neumann boundary conditions for solving both the $P$ and $\psi$ potentials, they are the same potential up to a constant scale factor $\psi=\Omega P$. Thus we get:
\begin{eqnarray}
\deriv{H_R^{NI}}{t} &=& 2 \Omega \int dA \ \left[(\partial_x P)^2 + (\partial_y P)^2\right] \geq 0
\end{eqnarray}
We can immediately see that the for any non-zero $\Omega$ we always get a positive contribution to the relative helicity injection from the non-inductive component of the electric field. On the other hand, when we use Assumption \ref{Eq-Ass2} to estimate the non-inductive component the $\psi$ and $P$ potentials decouple and thus no immediate positive bias is introduced to the relative helicity injection. Indeed, Assumption \ref{Eq-Ass2} with the energy-optimized value of $U$ reproduces the time evolution of the relative helicity injection much better, having a similar temporal trend as the \citetalias{Kazachenko2015} and our DAVE4VM estimates (however, exhibiting a consistent underestimation, see Section \ref{S-Helicity_inj_results}). 

It is not a surprise that neither of the ad hoc assumptions \ref{Eq-Ass1} or \ref{Eq-Ass2} are able to reproduce the injection of both the magnetic energy and relative helicity accurately. Still, the results clearly show that Assumption \ref{Eq-Ass2} succeeds much better than Assumption \ref{Eq-Ass1} in this sense. This implies that Assumption \ref{Eq-Ass2} is more suitable for practical use, for example when the electric fields are used as a boundary condition in coronal modeling. On the other hand, Assumption \ref{Eq-Ass1} offers an interesting way to study the effect of overestimating the photospheric helicity injection while still producing approximately correct energy injection in data-driven (TMF) simulations. For example, a comparative TMF simulation study using electric field boundary conditions based on (optimized) Assumptions 1 and 2 would be revealing in this sense. 

Finally, we study also the effect of some of the data processing methods on the results, in particular on the magnetic energy injection. These include the masking threshold used in removal of noise-dominated pixels from the data and the tracking speed of the active region, both of which have so far been poorly covered in the literature. We test three masking thresholds that correspond to approximately 1.5, 2.5 and 3 times the nominal noise level of the total magnetic field $B$ in SDO/HMI vector magnetograms (100 Mx cm$^{-2}$), from which the second one was the default used in this paper. We find that the lower the threshold the larger the magnetic energy and relative helicity injections are. The changes introduced to the energy injection by varying the mask are within the upper error bars of our reference energy injection estimates, but clearly above the lower error estimate mainly considered in this paper. Thus, if one wishes to reach a consistency with \citetalias{Kazachenko2015} (or with some other similar reference estimate of the energy injection) approximately within the $\pm$14\% error bounds (as we do) the mask threshold should be chosen consistently with the reference. However, the choice of the mask does not affect the main findings made above. The DAVE4VM estimate for the energy injection is completely insensitive to the choice of mask, even when the mask is removed entirely, implying that DAVE4VM cannot retrieve any significant vertical Poynting flux signal from the noise-dominated weak field pixels. Energy estimates based on ELECTRICIT inversions, in turn, increase significantly when the mask is removed. The fact that this increase is only seen in the ELECTRICIT output but not in the DAVE4VM inversions implies that this increase is a spurious result arising from the method-related properties of ELECTRICIT inversion, thus emphasizing the need for using a mask. 

For the tracking speed of the active region we tested three alternatives and found that our three ELECTRICIT estimates are insensitive to the choice of the tracking speed. However, the tracking speed had a more considerable effect on the DAVE4VM estimate, particularly when the tracking of the active region in vector magnetograms was reinforced using additional coalignment that minimized the whole-frame-shifts between successive frames. However, the change was still within the previously reported error bars of the DAVE4VM inversion, and the resulting DAVE4VM energy injection estimate was still most of the time within the lower error bars of \citetalias{Kazachenko2015} estimate discussed above. The stronger effect on DAVE4VM likely arises from the fact that the method explicitly tracks the image motions in vector magnetograms, which is more sensitive to the tracking speed and to the related changes in the velocity of the rest frame. Overall, despite the considerable effects that we found when varying the mask threshold and the tracking speed, none of the effects were sufficiently large to conclusively allow preferring some of the options over others.

\section{Conclusions}

By studying the area- and time-integrated Poynting flux we are able to show that the data processing and electric field inversion approach presented in this paper, which employs only photospheric magnetic field estimates is capable of reproducing the time evolution of the total -- and implicitly also the free -- magnetic energy injected through the photosphere, when compared to recently published state-of-the-art estimates. The lack of the photospheric velocity estimates in the approach is compensated using ad hoc assumptions whose free parameters allow optimization of the magnetic energy injection. Despite the good reproduction of the magnetic energy injection the ad hoc assumptions reproduce the injection of relative helicity poorly reaching at best a modest underestimation of the reference values from previous studies. The ad hoc assumption that constrains the non-inductive potential by setting the horizontal divergence of the electric field proportional to the vertical current density performs the best in this sense, implying that it is the most suitable assumption for practical applications. Finally, we show that our results are not very sensitive to the choice of the masking threshold for removing the noise-dominated pixels and the tracking speed of the active region. However, we find that our inversion requires some mask for the weak field pixels, and that this mask should be chosen consistently with the reference when optimizing the free parameters of the inversion using the magnetic energy injection. The main datasets used in this paper can be found from \url{https://zenodo.org/record/1034404}.

\begin{acks}
We thank the HMI team for providing us with the LOS and vector magnetic field SDO/HMI data. EL acknowledges the doctoral programme in particle physics and universe sciences (PAPU) of the University of Helsinki and the Magnus Ehrnrooth Foundation for financial support. EK and JP acknowledge the European Research Council (ERC) under the European Union's Horizon 2020 Research and Innovation Programme Project SolMAG 4100103. This research has made use of SunPy, an open-source and free community-developed solar data analysis package written in Python \citep{Mumford2015}. This is a pre-print of an article published in Solar Physics. The final authenticated version is available online at: \url{https://doi.org/10.1007/s11207-017-1214-0}.
\end{acks}

\begin{flushleft}
\scriptsize 
\textbf{Disclosure of Potential Conflicts of Interest} The authors declare that they have no conflicts of interest.
\normalsize
\end{flushleft}

\bibliographystyle{spr-mp-sola}
\bibliography{refs}  

\appendix
\section{Modifications to the existing data processing methods}
	\label{S-Appendix-mods_to_mets}

\subsection{Tracking scheme of active regions in SHARP data}
	\label{S-Appendix-Mods_to_SHARP_tracking}
	
As explained by \citet{Bobra2014} and \citet{Hoeksema2014} re-projected vector magnetograms in SHARP time series are created by tracking each HMI Active Region Patch (HARP) using a fixed rotation rate for the center point of the patch. The rotation rate is determined from the latitude of the center point using a differential rotation profile. Instead of the differential rotation profile described in \citet{Hoeksema2014} we employ the ''2 day lag'' profile from \citet{Snodgrass1983}. For the NOAA 11158 vector magnetogram time series used in this paper our tracking speed is $\sim$2 km\,s$^{-1}$ (0.27 deg per day) slower than the SHARP tracking speed. We observe that when using this method the center of the active region remains slightly better fixed to the center of the patch. As explained by \citet{Sun2013} the final re-projected SHARP magnetogram is created for each frame so that the disambiguated full-disk vector magnetogram is re-projected on the solar surface, \emph{i.e.}, interpolated from image pixels to a new grid that corresponds to Lambert CEA projection of the solar surface covering the predescribed active region size. The CEA projection is done so that the active region is ''viewed directly from above'', \emph{i.e.} using a heliographic coordinate system where the center of the active region is at the intersection of the central meridian and the equator. The grid spacing is 0.03$^{\circ}$ in projected heliographic coordinates corresponding to the SDO/HMI resolution at the disk center. We have modified this procedure so that instead of a CEA projection we employ Mercator projection, since conformal mappings are preferred when optical flow methods (such as DAVE4VM) are used \citep{Welsch2009,Kazachenko2015}. After re-projecting the full-disk magnetogram data we rotate the interpolated magnetic field vectors from SDO/HMI image basis $(B_{\xi},B_{\eta},B_{\zeta})$ (uniquely defined by the orthogonal image axes and LOS direction) to the heliographic basis $(B_r,B_{\theta},B_{\phi})$. The heliographic basis is chosen consistently with the Mercator map projection so that the center point of the active region is at the intersection of the central meridian and the equator. Transformation from the default heliographic basis $(B_r,B_{\theta},B_{\phi})$ where the true solar equator has latitude $\lambda = 0$ to the patch-centered basis $(B_r',B_{\theta}',B_{\phi}')$ can be be done using the following transformation (which is not included in the references listed above):
\begin{eqnarray}
\label{Eq-rotate_vec_to_patch_centered}
\left(\begin{array}{c} B_{r}' \\ B_{\theta}' \\ B_{\phi}' \end{array}\right) &=& \left(\begin{array}{ccc} 1 & 0 & 0 \\ 0 & \cos \alpha & -\sin \alpha \\ 0 & \sin \alpha & \cos \alpha \end{array}\right) \left(\begin{array}{c} B_{r} \\ B_{\theta} \\ B_{\phi} \end{array}\right) \\
\label{Eq-rotate_vec_to_patch_centered_alpha}
\alpha(\lambda,\phi,\lambda_c,\phi_c) &=& \textrm{sgn}(\lambda_c)\textrm{sgn}(\phi - \phi_c) \times \nonumber \\
&& \cos^{-1} \left[ \left(\cos \lambda_c \cos \lambda + \sin \lambda_c \sin \lambda \cos (\phi - \phi_c) \right)/ \cos \lambda' \right] \\
\label{Eq-rotate_vec_to_patch_centered_lambda}
\lambda' &=& \sin^{-1} \left[\sin \lambda \cos \lambda_c - \sin \lambda_c \cos \lambda \cos \left(\phi - \phi_c \right) \right],
\end{eqnarray}
where $\lambda'$ is the latitude in heliographic coordinates where the patch center $(\lambda_c',\phi_c') = (0,0)$.

This is different from SHARP data in which the heliographic basis corresponds to Carrington coordinates in which the equator is the true solar equator \citep{Sun2013}. Finally, we further transform into local Cartesian basis where we treat the projected active region patch as a flat surface:
\begin{eqnarray}
\left(\begin{array}{c} B_x \\ B_y \\ B_z \end{array}\right)  &=& \left(\begin{array}{c} B_{\phi}' \\ -B_{\theta}' \\ B_r', \end{array}\right)
\end{eqnarray}
where we drop the apostrophes to simplify notation.

\subsection{Algorithm for removal of spurious flips in azimuth of the magnetic field}
	\label{S-Appendix-Mods_disambig_flip_removal}	
	
\citet{Welsch2013} define spurious flips of the azimuth of the magnetic field $\phi$ as sudden jumps where $\phi$ jumps approximately from one disambiguation to the other $\phi \rightarrow \phi + 180^{\circ}$ and then quickly back. In a time series this flipping can be seen as rapid blinking in the transverse magnetic field components. The authors proposed a temporal smoothing algorithm to recognize and remove such spurious flips. The algorithm detects these flips using two conditions. First, the change in azimuth between successive frames must be significant enough (over $120^{\circ}$):
\begin{equation} \label{Eq-cond1_for_flip}
||\phi(x_m,y_n,t_k) - \phi(x_m,y_n,t_{k-1})|| > 120^{\circ}.
\end{equation}
where $||x||$ refers to the absolute acute angle of $x$:
\begin{equation} \label{Eq-acute_angle_def}
||x|| = \left\lbrace \begin{array}{c} |x|, |x| \leq 180^{\circ} \\
				360^{\circ} - |x|, |x| > 180^{\circ}
		 \end{array} \right.
\end{equation}
Second, if a spurious azimuth flip candidate at $t_k$ is removed by flipping the azimuth back to the ''correct'' value ($\phi \rightarrow \phi + 180^{\circ}$), the procedure must decrease the sum of unsigned azimuth differences over the nearby frames:
\begin{equation} \label{Eq-cond2_for_flip}
S = \sum_{l = -R}^{R} ||\phi(x_m,y_n,t_{k+l}) - \phi(x_m,y_n,t_{k})||,
\end{equation}
\emph{i.e.}, the procedure must smooth the azimuth in the time domain.
Here $R$ is the expected length of that spurious flip in time, and it gives the upper limit for the number of frames the spurious flip may last in order to be recognized by the algorithm. $R$ is the only free parameter in the algorithm and \citet{Welsch2013} suggested two values for it: $R=2$ or $R=4$, from which they used the former. We have chosen to use $R=4$ instead, since it produces more stable evolution of the azimuth in visual inspection. Another major qualitative departure from the work of \citet{Welsch2013} is that we applied the flip removal procedure after interpolating the SDO/HMI data to local Cartesian system in Mercator projection (Appendix \ref{S-Appendix-Mods_to_SHARP_tracking}), whereas they applied the procedure in native HMI pixels. Thus, in our approach the erroneous values of pixels where the azimuth had a spurious flip propagated in the interpolation. Due to this we employ an additional step after recognizing and fixing the spurious flips: we smooth each fixed pixel and all of its 8 neighboring pixels using a Gaussian smoother with $\sigma = 1$ pixel truncated at $1\sigma$ as a way to mitigate the effect of propagating the erroneous azimuth to the neighboring pixels in the interpolation. An additional motivation for this smoothing is that it removes some of the artifacts of the flip removal procedure itself: the procedure leaves some ragged structures and singular pixels of inconsistent azimuth values to the data, which we consider spurious, particularly when taking into account that the minimum energy disambiguation tries to reduce currents, \emph{i.e.} to smooth the azimuth spatially.

\section{Numerical implementation of the electric field inversion methods in ELECTRICIT}
	\label{S-Appendix-num_impl_of_E_inv}	
In the PDFI method Faraday's law (Eq. \ref{Eq-F_law_ind}) is uncurled using the Poloidal-Toroidal Depomposition of the magnetic field \citep{Fisher2010}:
	\begin{eqnarray} 
		\vec{A} &=& \curl{P\uvec{z}} + T\uvec{z} \label{Eq-PTD_A} \\
		\vec{E}_I &=& -\curl{\dot{P}\uvec{z}} - \dot{T}\uvec{z}, \label{Eq-PTD_E_I}
	\end{eqnarray}			
where $\vec{A}$ is the magnetic vector potential $\curl{\vec{A}} = \bb$, and $\dot{P}$ and $\dot{T}$ are the partial time derivatives of two-dimensional poloidal and toroidal potentials, respectively. They can be solved from the Poisson equations of which source terms are determined by the $\partial \vec{B}/\partial t$ field. The time derivative of the poloidal potential can be solved from:
	\begin{equation}		\label{Eq-Poisson_for_P}
		\nabla_h^2 \dot{P} = -\pder{B_z}{t},
	\end{equation} 
where $\nabla_h^2$ is the horizontal Laplacian $\partial_x^2 + \partial_y^2$. In this paper we employ only the horizontal components of the inductive electric field $\vec{E}_h^I = (E_x^I,E_y^I)$ which are determined solely from $\dot{P}$. We solve the Equation \ref{Eq-Poisson_for_P} using the same homogeneous Neumann boundary conditions and the numerical solver (FISHPACK, \citealp{Swarztrauber1975}) as \citet{Kazachenko2014}. As noted by \citet{Kazachenko2014}, FISHPACK modifies the source term of a Poisson equation by adding a constant value to it, if it is found to be inconsistent with the homogeneous Neumann boundary conditions. As a result of this correction the output electric field is slightly inconsistent with Faraday's law. To remove this inconsistency we add post facto correction to the horizontal electric field components as done by \citet{Fisher2010}. To further ensure the consistency with Faraday's law we have also decided to use different finite difference approximation for Equations \ref{Eq-PTD_A}  and \ref{Eq-PTD_E_I} than \citet{Kazachenko2014}. The FISHPACK solver employs the 5-point stencil for the Laplacian, which implicitly assumes that the first order spatial derivatives of the solution of the Poisson equation are determined in half-grid points. Thus, if the first order derivatives in Equation \ref{Eq-PTD_E_I} and Faraday's law (Eq. \ref{Eq-F_law_ind}) for the horizontal components of $\vec{E}_I$ are calculated using a typical central difference scheme, $\partial B_z/\partial t$ is not reproduced exactly. \citet{Kazachenko2014} argue that the error introduced by this is small and employ a central difference scheme also for the first order derivatives. However, we wish to reproduce $\partial B_z/\partial t$ from Faraday's law exactly, and therefore employ the following consistent finite difference formulas for the spatial derivatives (used in a similar fashion also by \citealp{Yeates2017}):
\begin{eqnarray}
E_x^I(x_m,y_{n+1/2}) &=& -\frac{\partial \dot{P}}{\partial y}(x_m,y_{n+1/2}) = -\frac{\dot{P}(x_m,y_{n+1}) - \dot{P}(x_m,y_n)}{\Delta y} \label{Eq-ExI_from_P} \\
E_y^I(x_{m+1/2},y_n) &=& \frac{\partial \dot{P}}{\partial x}(x_{m+1/2},y_n) = \frac{\dot{P}(x_{m+1},y_n) - \dot{P}(x_m,y_n)}{\Delta x} \label{Eq-EyI_from_P} \\
\frac{\partial B_z(x_m,y_n)}{\partial t} &=& -(\nabla \times \mathbf{E}_I)_z(x_m,y_n) = \frac{\partial E_x^I}{\partial y}(x_m,y_n) - \frac{\partial E_y^I}{\partial x}(x_m,y_n) \label{Eq-dBzdt_from_E} \\ 
&=& \frac{E_x^I(x_m,y_{n+1/2}) - E_x^I(x_m,y_{n-1/2})}{\Delta y} \nonumber \\ && - \frac{E_y^I(x_{m+1/2},y_n) - E_y^I(x_{m-1/2},y_n)}{\Delta x}. \nonumber
\end{eqnarray} 
As one can implicitly read from the equations above the spatial derivatives consistent with the 5-point stencil give the electric field components in a staggered grid (Yee mesh,  \citealp{Yee1966}) with respect to $B_z$ and $\partial B_z/\partial t$ at $(x_m,y_n)$. Since many of the TMF simulations employ such a staggered grid (\emph{e.g.}, \citealp{vanBallegooijen2000,Cheung2012}), this makes the electric field inversions of ELECTRICIT instantly suitable to be used as the boundary condition data of such simulations, while being simultaneously consistent with the observed time evolution of $B_z$.

The numerical approach described above is used to solve also the $P$ potential from $B_z$ and the resulting horizontal components of the vector potential $\ap^h$ (see \citealp{Fisher2010} for details). Thus, our $\ap^h$ is determined in a staggered grid being also exactly consistent with $B_z$.

The non-inductive component $\psi$ is solved similarly as the $\dot{P}$ potential in Equation \ref{Eq-Poisson_for_P} (including the post facto corrections) for each Assumption 1 and 2 (Section \ref{S-E-Inv}). When calculating the current density $j_z$ for Assumption \ref{Eq-Ass2} the spatial derivatives of $B_x$ and $B_y$ are calculated using a central difference scheme, since we wish to keep $j_z$ and thereby also $\psi$ cospatial with the magnetogram data. However, since $\vec{E}_I$ is defined on a staggered grid the gradients $\grad_x{\psi}$ and $\grad_y{\psi}$ are not cospatial with $E_x^I$ and $E_y^I$. This can be fixed by interpolating $\psi(x_m,y_n)$ to $\psi(x_{m+1/2},y_{n+1/2})$, after which the gradients:
\begin{eqnarray}
\grad_x{\psi}(x_m,y_{n+1/2}) &=& \frac{\psi(x_{m+1/2},y_{n+1/2}) - \psi(x_{m-1/2},y_{n+1/2})}{\Delta x} \label{Eq-grad_x_psi} \\
\grad_y{\psi}(x_{m+1/2},y_n) &=& \frac{\psi(x_{m+1/2},y_{n+1/2}) - \psi(x_{m+1/2},y_{n-1/2})}{\Delta y}, \label{Eq-grad_y_psi}
\end{eqnarray}
are cospatial with $E_x^I$ and $E_y^I$ in Equations \ref{Eq-ExI_from_P} and \ref{Eq-EyI_from_P}. Moreover, this approach ensures that $(\curl{\vec{E}})_z$ in Faraday's law is unaffected by $-\nabla_h \psi$. 

Finally, it should be noted that the use of staggered grid in the ELECTRICIT solutions of $\vec{E}$ and $\ap$ requires also (linear) interpolation back to a centered grid whenever Poynting or relative helicity fluxes are calculated.

\end{article} 

\end{document}